\pdfoutput=1
\documentclass[twocolumn,showpacs,twoside,10pt,prl]{revtex4}

\usepackage{graphicx}
\usepackage{epsfig}
\usepackage{epsf}
\usepackage{amssymb}
\usepackage{amsmath}
\usepackage{amsthm}
\usepackage{multirow}
\usepackage[usenames]{color}
\definecolor{Red}{rgb}{1,0,0}
\definecolor{Blue}{rgb}{0,0,1}


\begin{document}

\title{Implementation of a Quantum-simulation Algorithm of Calculating Molecular Ground-state Energy on an NMR Quantum Computer}
\author{Jiangfeng Du$^{\ast}$, Nanyang Xu, Xinhua Peng, Pengfei Wang, Sanfeng Wu, Dawei Lu \\
\normalsize{Hefei National Laboratory for Physical Sciences at
Microscale and Department of Modern
Physics,University of Science and Technology of China, Hefei, Anhui 230026, People's Republic of China}\\
\normalsize{$^\ast$To whom correspondence should be addressed;
E-mail:  djf@ustc.edu.cn}}

\begin{abstract}
It is exponentially hard to simulate quantum systems by classical algorithms, while quantum computer could in principle solve this problem polynomially. We demonstrate such an quantum-simulation algorithm on our NMR system to simulate an $H_2$ molecule and calculate its ground-state energy. We utilize the NMR interferometry method to measure the phase shift and iterate the process to get a high precision. Finally we get 17 precise bits of the energy value, and we also analyze the source of the error in the simulation.

\end{abstract}
\maketitle

\emph{Introduction}. --- It is well known that quantum algorithms provides an excellent speedup towards classical algorithms in many problems\cite{qcqi}. Among these, the most famous example is the Shor's algorithm\cite{shor_algorithm}, which solves the factoring problem exponentially faster than classical methods. Besides the success of quantum computation(QC) applied in computational problems, another important application of QC is the simulation of quantum systems, an idea conceived by Feynman\cite{feynman}. Simulation of realistic quantum systems requires an exponential amount of resources to handle the large number of coupled $Sch\ddot{o}dinger$ equations used to describe the particles in the system. However, within the QC architecture it was shown that universally simulating an physical system is efficient\cite{lloyld_univsim}, \emph{i.e.}, costing polynomial resources with the size of the target system.

Quantum simulation allows us to examine several new frameworks such as the many-particle system\cite{zalka1996,manybody}, pairing-Hamiltonian systems\cite{wu2002} and \emph{etc.}\cite{peng2005,oscillator,smirnov2007}. Specifically, several quantum-simulation algorithms were proposed to determine properties like thermal rate constant\cite{lidar1999} and molecular energies\cite{molenergy} in quantum chemistry. The calculation of molecular energies is a fundamental problem focused in computational quantum chemistry. On classical computer, resources for a full simulation of the molecular system scale exponentially with the number of atoms involved, limiting such full configuration interaction calculations of molecular energies to diatomic and triatomic molecules\cite{chemphys}. However the calculation could be efficient if it is performed on a quantum simulator using the algorithm proposed by Aspuru-Guzik \emph{et al} \cite{molenergy}, where information about the energy is transferred onto the phase shift of a quantum register and measured by a quantum phase estimate algorithm (PEA) iteratively.

Because of the practical importance of QS, experimental demonstrations of these algorithms are not only of fundamental interests, but also a cornerstone of emerging field of new technologies. Previous experiments have been performed to simulate some small systems on NMR\cite{oscillator,negrevergne,peng2005} or ion-trap\cite{qumagnet} platforms. However no such demonstrations have been performed till now due to the requirement of controlling sufficient number of qubits for simulation a molecule in original Aspuru-Guzik's algorithm.

In this letter, we adopt two methods to overcome the challenges: 1) choosing simplest
case of the smallest molecule, \emph{i.e.} $H_2$; 2) utilizing the
mature technology on NMR platform - the NMR interferometer - to
measure the phase shift. We also adopt a improved iterative scheme
to extend the precision to a high level. By these methods, we
realize this algorithm and calculate the ground-state energy
of $H_2$ molecule to a precision of 17 bits. We also find that the
precision of the result got from this algorithm is limited by the
imperfection of simulating the operators in experiment.

\emph{The Hamiltonian of $H_2$ molecule}. --- Limited by the current
technology, to calculate a large molecular energy in Full
Configuration Interaction is not possible, so we choose the simplest
situation: the ground-state energy of Hydrogen molecule in the
minimal STO-3G basis\cite{quantchemistry}. The electron's
Hamiltonian of $H_2$ molecule with Born-Oppenheimer approximation is
shown as follows\cite{quantchemistry} ,
$$
    H=\sum^2_{i=1}(T_i+\sum^2_{j=1}V_{ij})+\sum^2_{\substack{i,j=1\\i>j}}O_{ij}
$$
where $T_i$ is the kinetic energy of the $i$th electron, and
$V_{ij}$ is the coulomb potential energy between the $i$th electron
and the $j$th nucleus, while $O_{ij}$ is the coulomb potential
energy between the $i$th and $j$th electron. In this two-nucleus and
two-electron molecule in STO-3G basis, each atom has a $1s$
Gaussian-type function, and the two $1s$ functions compose one
bonding orbital with gerade symmetry and one antibonding orbital
with ungerade symmetry. So there are 4 spin orbitals which can form
6 configurations. Considering the singlet symmetry and the spatial
symmetry of $H_2$ exact ground state, only two configurations are
acting in fact in the calculation: the ground state configuration
$|\Psi_0\rangle$ and the double excitation configuration
$|\Psi_{1\bar{1}}^{2\bar{2}}\rangle$. Thus, the Hamiltonian matrix
is(in atom units, the nucleus distance is 1.4\emph{a.u.}, only the
electron's energy):

\begin{eqnarray}
        H&=&
        \begin{pmatrix}
            \langle\Psi_0|H|\Psi_0\rangle & \langle\Psi_{1\bar{1}}^{2\bar{2}}|H|\Psi_{1\bar{1}}^{2\bar{2}}\rangle\\
            \langle\Psi_{1\bar{1}}^{2\bar{2}}|H|\Psi_0\rangle & \langle\Psi_{1\bar{1}}^{2\bar{2}}|H|\Psi_{1\bar{1}}^{2\bar{2}}\rangle
            \end{pmatrix}\nonumber\\
        &=& \begin{pmatrix}
            -1.8310 & 0.1813\\
            0.1813 & -0.2537
            \end{pmatrix}
\end{eqnarray}
whose theoretical eigenvalue is -1.8516 \emph{a.u.}.

\begin{figure}[htb]
\begin{center}
\includegraphics[width= 0.99\columnwidth]{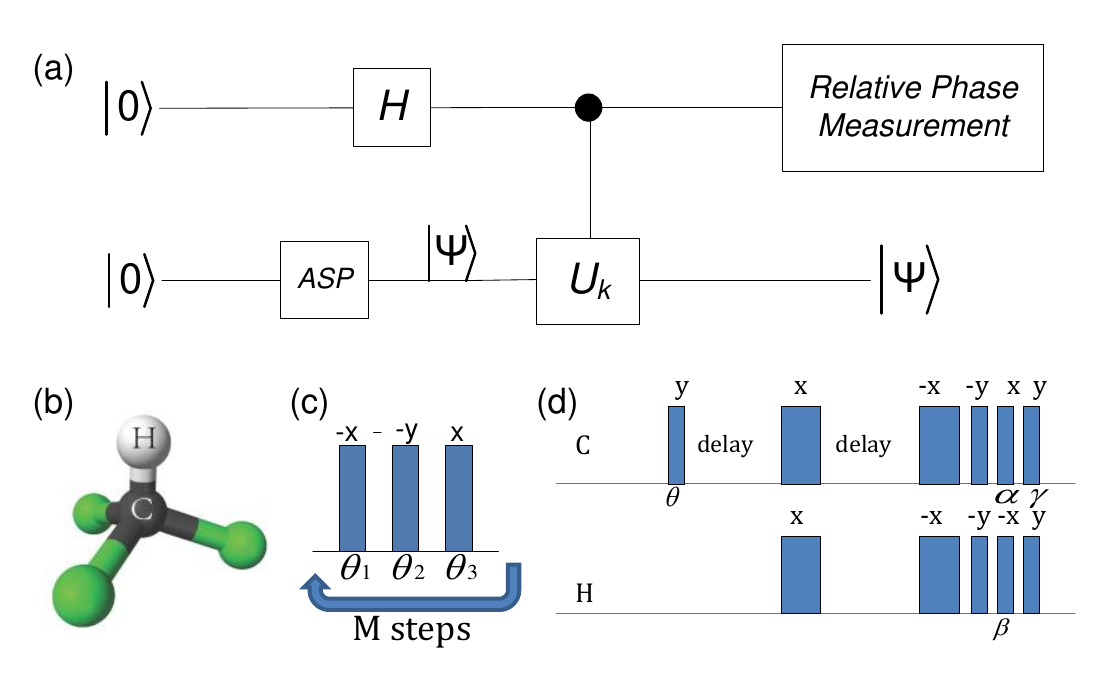}
\end{center}
\caption{(a) General schematic circuit for calculating molecular
energies. (b) Molecular structure of quantum register ($CHCl_{3}$).
(c) Pulse sequence for the adiabatic process to prepare the $^{13}C$
nucleus in the ground state $|\Psi\rangle$ of the molecular
Hamiltonian $H$.(d) Pulse sequence to implement the controlled-$U_k$
operation,where $\theta=0.226$, $\gamma=1.3458$,
$\beta=\beta_{k}^{0}-\phi^{'}_{k-1}(\phi^{'}_{k-1}$ is calculated
from the result of last iteration and
$\beta_{k}^{0}=[2.0233,-2.6629,-2.4539,-0.7814,0.0322,0.2575])$ for
the k$th$ iteration, and for the first control U operation,
$\alpha=\frac{\pi}{2}$ and $delay=\frac{1}{4J_{wa}}$ while
$\alpha=0$,$delay=0$ for other 5 times iteration. In fact, we could
simplify the sequence to only three pulses on qubit $^{1}H$ except
the first control $U$ operation. The $U$ sequences are optimized as
we chose a suitable $\tau=1.941122$.}\label{circuit_exp}
\end{figure}

\emph{The calculation of molecular energy}. --- As shown in
Fig.\ref{circuit_exp}a, the calculation of the molecular energy in
our experiment is described by four steps: 1) adiabatic preparation of the
system qubit to the ground state of the Hamiltonian $H$ 2) application of
the time evolution of the molecular Hamiltonian on the qubits to
generate the phase shift on the probe qubit; 3) measurement of the phase shift on the probe qubit to extract the energy information. We will
introduce the three steps in detail as follows.

In the first step, the quantum simulator's system qubit is prepared by
Adiabatic State Preparation (ASP) to $|\Psi\rangle$, the ground
state of the molecular Hamiltonian $H$. According to the
\emph{quantum adiabatic theorem}\cite{Messiah,Kato}, the qubit is
prepared on an simple Hamiltonian's ground state and the systematic
Hamiltonian of the qubit varies slowly enough from the simple
Hamiltonian to $H$, if there's an energy gap between the ground
state and the first excited state, the qubit will stay on the
instantaneous ground state of the system Hamiltonian. Thus the qubit
finally is prepared on the ground state of $H$ after the ASP.

In the second step, An unitary operator $U=e^{-iH\tau}$ is
applied to the state $|\Psi\rangle$, with only generating a phase
shift to the probe qubit by the controlled operation. Here
$U|\Psi\rangle=e^{-iH\tau}|\Psi\rangle=e^{i2\pi\phi}|\Psi\rangle$
where $E=-2\pi\phi/\tau$ is the energy of $H$'s ground state. Note
that the energy $E$ is negative so we make the phase $\phi$ to be
positive and $\tau$ is chosen properly to make the phase $\phi$
ranges from 0 to 1.

Finally in the measurement step,  a four-bit inverse
Quantum Fourier Transform (QFT) is adopted as the Relative Phase
Measurement to evaluate the phase shift in the Aspuru-Guzik's
proposal. This apparatus needs four qubits as probe qubits to get one
precise bit with successful possibility of $15/16$\cite{qcqi}. While
in the NMR platform there's a mature technology NMR interferometer,
named from the similar apparatus originally used in optics, which
could easily measure the relative phase shift of the quantum
states by
modulating the spectrum patterns\cite{du_geophase,peng_comple}. On our NMR interferometer, the
phase shift could be evaluated with an error bound of less than
$\pm5^\circ$, much more precise than the performance of the original
four-bit inverse QFT apparatus. Thus we utilize the interferometry
to measure the phase shift in our experiment.

\emph{The iterative scheme}. --- For useful practical application, it should be possible to iterate the above process to achieve arbitrary precision in the molecular energy. We made a small modification to the iterative scheme
in Aspuru-Guzik's algorithm to improve its reliability. As shown in Fig.\ref{circuit_exp}a, for each iteration $k$ we applied the controlled-$U_k$ and measure the phase shift. We start the iterations
from $U_0=U$ and iterate the process by choosing $U_{k+1}=[e^{-i2\pi\phi^{'}_k}U_k]^{2^n}$.
Here, $n$ is the number of bit attained in each iteration and $\phi^{'}_k = max\{\phi_k-\phi_{errbd},0\}$ where $\phi_k$ is the phase shift measured in the $k$th iteration. Note that $n$ is limited by the precision of the phase measurement in each iteration, \emph{i.e.}, $2^{-n}\geq 2\phi_{errbd}$.

\emph{The experimental implementation.} --- We used the
$^{13}C$-labeled chloroform dissolved in $d_6$-acetone as a
two-qubit NMR quantum computer, where $^{13}C$ nucleus was used for
the system qubit while $^{1}H$ is for the probe qubit. The
molecular structure is shown in Fig.\ref{circuit_exp}(b). The
natural Hamiltonian of this two-qubit system is given by:
\begin{eqnarray}\label{HamCHF}
\mathcal{H}_{\mathit{NMR}}=\frac{\omega_{p}}{2}\sigma_{z}^p+\frac{\omega_{s}}{2}\sigma_{z}^{s}+\frac{\pi
J_{ps}}{2}\sigma_{z}^{p}\sigma_{z}^{s}
\end{eqnarray}
where $\omega_{s}/2\pi$ and $\omega_{p}/2\pi$ are the Larmor
frequencies of nucleus $^{13}C$ and $^{1}H$, and $J_{ps}$ represents
the J coupling constant, typically, $J_{ps}=214.6Hz$. The
experiments were carried out at room temperature on a Bruker AV-400
spectrometer. Let us now describe the experiment in detail.

A) Preparation of the initial state: Starting from the thermal
equilibrium state, we first created a pseudo-pure state (PPS)
$\rho_{00} =\frac{1 - \epsilon}{4} \mathbf{I}+ \epsilon|\uparrow
\uparrow \rangle \langle \uparrow \uparrow |$ using the spatial
average technique\cite{Cory:1997aa}, with $\mathbf{I}$ representing the
$4 \times 4$ unity operator and $\epsilon \approx 10^{-5}$ the
polarization. However, the simulation algorithm requires the system
qubit is in the ground state $\vert \psi_g \rangle$ of the
Hamiltonian of the $H_2$ molecule [Eq. (2)] while the probe bit
is in the $\vert + \rangle = \frac{1}{\sqrt{2}} \vert \uparrow
\rangle + \vert \downarrow \rangle$. The $\vert + \rangle$ state was
easily prepared by a pseudo-Hadamard gate $R_{y}^{H} (\pi/2)$ from
$\vert \uparrow \rangle$. On the other hand, the unknown ground
state $\vert \psi_g \rangle$ can be prepared by an adiabatic
procedure where we starts from a simple initial Hamiltonian $H_0$
whose ground state $\vert \psi_0 \rangle$ is easily constructed. The
adiabatic theorem tells that if the system Hamiltonian $H_{ad}(t)$
varies slowly enough so that the adiabatic condition is satisfied
\cite{Messiah}, the system remains in its instantaneous ground
state. Therefore, if $H_{ad}(t)$ reaches the Hamiltonian $H$ at $t=
T$,  the system is prepared the ground state $\vert \psi_g \rangle$
of $H$. The time-dependent Hamiltonian during the adiabatic passage
is obtained by linear interpolation: $H_{ad}=(1-s)\sigma_x + sH$
with $s=\frac{t}{T}$. Here we chose the initial Hamiltonian $H_0 =
\sigma_x$ whose ground state is $\vert - \rangle =
\frac{1}{\sqrt{2}} \vert \uparrow \rangle - \vert \downarrow
\rangle$ prepared by the conjugated pseudo-Hadamard gate $R_{y}^{H}
(-\pi/2)$. We implemented the adiabatic preparation by discretizing
the continuous adiabatic passage with the optimized parameters
\cite{Peng_factor,Steffen}: the discrete steps $M+1 = 6$ and the
total time $T=5.4ms$. Therefore the unitary evolution for the
discrete adiabatic passage is then
$U_{ad}=\prod_{m=0}^{M}U^{ad}_{m}=\prod_{m=0}^{M} e^{-iH_m\tau },$
where the duration of each step is $\delta =T/(M+1)$. For each step,
\begin{equation}
U^{ad}_{m}=e^{-i\frac{\delta}{2}(1-s_m)\sigma_{x}}e^{-is_mH \delta}
e^{-i\frac{\tau}{2}(1-s_m)\sigma_{x} }+O(\delta ^{3}),
\end{equation}
where $s_m = \frac{m}{M+1} T$. Consequently,  $U^{ad}_{m}$ can be
implemented by the pulse sequence
$R_{-x}^{C}(\theta_{1})-R_{-y}^{C}(\theta_{2})-R_{x}^{C}(\theta_{3})$
shown in Fig.\ref{circuit_exp}(c). The fidelity of the state
obtained by the adiabatic preparation is around $98.93\%$ comparing
with the theoretical expectation 0.996.

B) The controlled-$U_k$ operation.  The controlled-$U_k$ has the
form of
$$
\mathcal{U}_k = | \uparrow \rangle\langle \uparrow |\otimes I +|
\downarrow\rangle\langle \downarrow |\otimes U_k.
$$
For the first iteration, \emph{i.e.}, $k=0$, $U_0 = e^{-iH\tau}$.
Thus, the $\mathcal{U}_0$ operation transforms the initial state
$\psi_{in} = \frac{1}{\sqrt{2}}(\vert \uparrow \rangle + \vert
\downarrow \rangle ) \vert \psi_g \rangle$ into $\psi_{f}
=\frac{1}{\sqrt{2}}(\vert \uparrow \rangle + e^{i2\pi \phi} \vert
\downarrow \rangle ) \vert \psi_g \rangle$ where $\phi = - E \tau /2
\pi$ with the energy $E$. As in an interferometer, the controlled
logic gate effectively introduces a relative phase shift $2\pi \phi$
between 'two paths': the $\vert 0 \rangle $ and $\vert 1 \rangle $
states in the initially prepared superposition of the auxiliary
qubit, which can be read out directly in NMR \cite{peng_comple,
du_geophase}. The value of $\tau$, in principle, can be arbitrarily
chosen to make $\phi \in (0, 1)$. For experimental convenience, we
chose
$\tau=\frac{\pi}{\sqrt{(2H(1,2))^2+(H(1,1)-H(2,2))^2}}=1.941122$,
and the pulse sequence to implement the controlled-$U_k$ operator is
shown in Fig.\ref{circuit_exp}(d). The different $\mathcal{U}_k$ in
each iteration is realized by adjusting the parameters $\alpha,
\beta$ and $delay$ .

\begin{figure}[htb]
\begin{center}
\includegraphics[width=0.99\columnwidth]{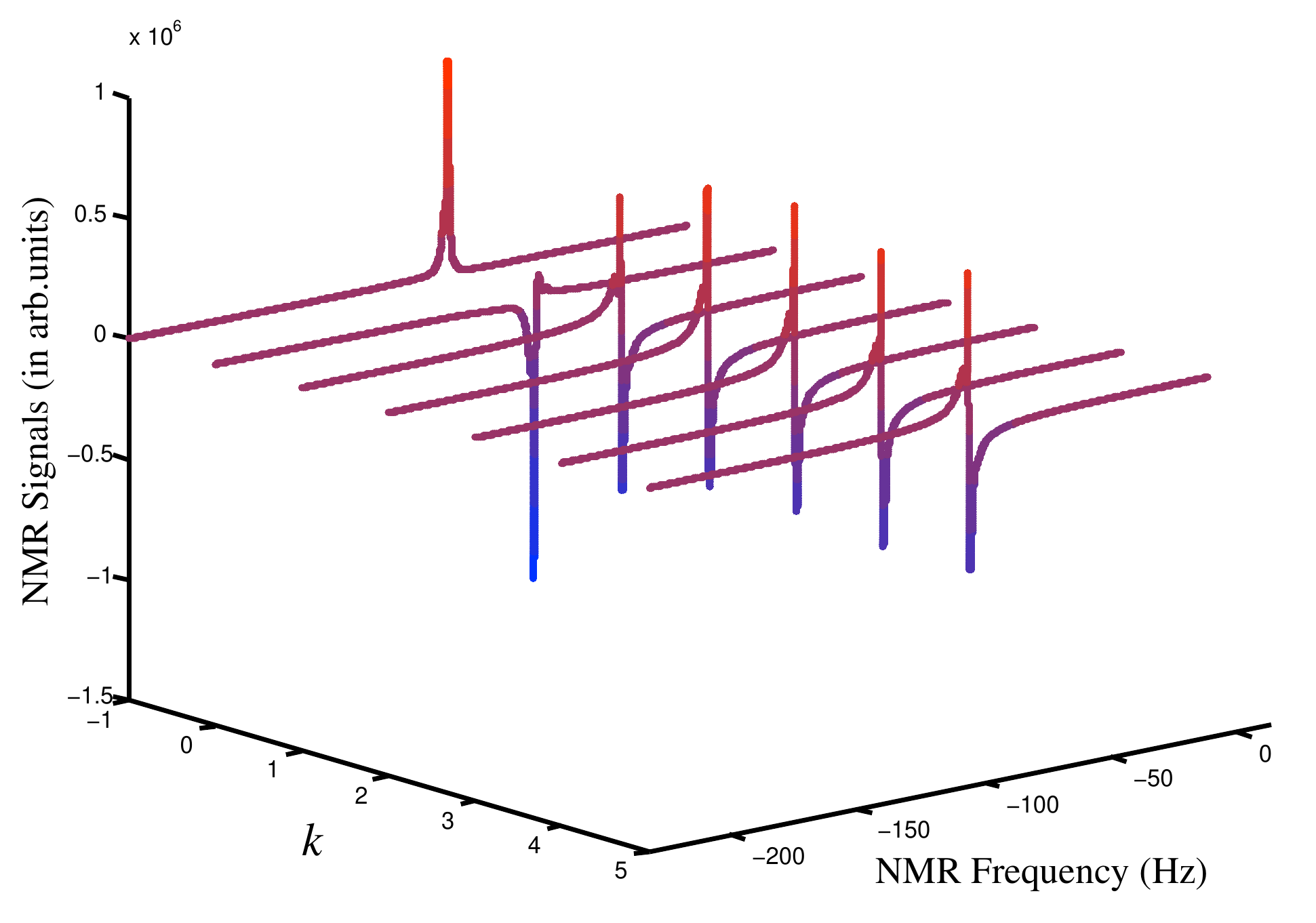}
\end{center}
\caption{Experimental $^{1}H$ spectra for all six iterations along
with the reference spectrum of the initial state $\psi_{in}$
(denoted by $k = -1$ as the dashed line). The following spectra
($k=0\sim5$) are observed after each iteration, which provides the
relative phase information. The phase measured in this iteration is
used in the next iteration.} \label{expspectrum}
\end{figure}

C) Measurement.  The relative phase shift is obtained if we measure
the NMR signal of the auxiliary qubit ($\langle \sigma^{-}_a
\rangle$):
$$
\langle \sigma^{-}_a \rangle = \langle \psi_f \vert \sigma^{-}_a
\vert \psi_f \rangle = \cos (2 \pi \phi) + i \sin (2 \pi \phi).
$$
As a result, the quadrature detection in NMR serves as a phase
detector, \emph{i.e.}, the Fourier-transformed spectrum gives the
relative phase information. Here we take the initial state of
$\psi_{in}$ as the reference phase. The experimental spectra for
each interaction are showed in Fig.\ref{expspectrum}.

After each time iteration of the above procedure, we measure the phase
shift and prepare the operator for the next iteration. After measured the
phases, we uses a recursive method to rebuild the $\phi$ as the
experiment result. The recursive method is formulated as follows
,for the result $\phi_{exp}$ after $k$th iteration,
$$
\phi^{c}_{i-1}=\phi^{c}_{i} / \phi_{errbd} + \phi^{'}_{i-1}.
$$
$\phi^{c}_{i}$ is the intermediate value only for calculation. The
recursive variable $i$ iterates from $i=k$ to $1$ with $\phi^{c}_k =
\phi_k$ where $\phi_k$ is the measured phase in the $k$th iteration.
And finally we get the result by $\phi_{exp}=\phi^{c}_0$.

\begin{figure}[thb]
\begin{center}
\includegraphics[width= 0.99\columnwidth]{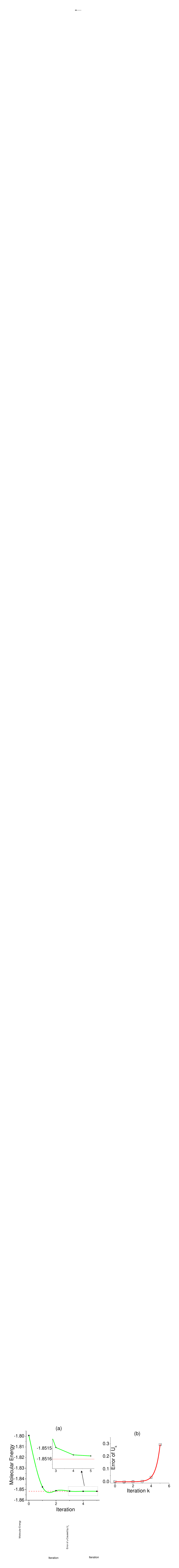}
\end{center}
\caption{(a)The energy values got from the experiment in each iteration. The green line is the spline fit of experimental values and the dash
red line is theoretical expectation. The experimental
value approaches theoretical expectation exponentially. The inside
figure shows the small gap between experimental value and theoretical expectation. (b) The error of phase shift generated by $U_k$ in the iterations. The error scales exponentially with $k$ due to the imperfection of the operator $U$.}\label{expresult}
\end{figure}

\begin{table}[htb]
\caption{\footnotesize The Theoretical
Expectation and Experimental result of $\phi$ after each Iteration. The experiments get 3
bits of $\phi$ in each iteration. The bits with an underline in each
$\phi_{exp}$ denotes the precision of the value. Overall, we get 17
bits of precision by 6 iterations.} \label{comparision}
\begin{center} {\footnotesize
\begin{tabular*}{0.48\textwidth}{@{\extracolsep{\fill}}|c|c|ccccc|}
\hline\hline
 & iteration & \multicolumn{4}{c}{binary values} &\\
\hline
 \multirow{6}{*}{$\phi_{exp}$}
 & 0 & 0.01\underline{\textcolor{Red}{0}}00 & 11100 & 10010 & 11000 & 10010 \\
 &  1 & 0.01001 & \underline{\textcolor{Red}{0}}0100 & 00111 & 01110 & 01001\\
 &  2 & 0.01001 & 001\underline{\textcolor{Red}{0}}0 & 11001 & 01010 & 11010\\
 &  3 & 0.01001 & 00100 & 1\underline{\textcolor{Red}{1}}011 & 10011 & 10001\\
 &  4 & 0.01001 & 00100 & 1101\underline{\textcolor{Red}{1}} & 11110 & 11100\\
 &  5 & 0.01001 & 00100 & 11100 & 00\underline{\textcolor{Red}{0}}00 & 01001\\
\hline
\multicolumn{2}{|c|}{$\phi_{th}$} & 0.01001 & 00100 & 11100 & 00101 & 01100  \\
\hline\hline
\end{tabular*} }
\end{center}

\end{table}

The result of the iteration is shown in Fig.\ref{expresult}a and Tab.\ref{comparision}. The
value $\phi_{exp}$ we got from experiment approaches the theoretical value
of $\phi_{th}$ rapidly as the iterations proceeds. However, there's still a gap between $\phi_{exp}$ and $\phi_{th}$ which is larger than $2^{-18}$ after 6 iterations. This means that although we get 18 bits from the experiment, only 17 bits of them are precise.

This is because the time evolution of molecular Hamiltonian $H$ is simulated by $U=e^{-iH\tau}$ in the
experiment and this simulation could not be arbitrarily precise.
Thus the error of $U_k$, which contains $2^k$'s power of $U$, scales exponentially with $k$.  We figure out this error in Fig.\ref{expresult}b, which shows that in the $5$th iteration the error is large enough and we can not get any more precise bits. Finally the molecular energy got from our experiment is -1.851569, with 17 precise bits towards the theoretical value -1.8516.

\emph{Conclusions.} --- We have demonstrated a quantum algorithm
to calculate the molecular energies on an 2-bit quantum computer.
This is one of early researches of quantum-simulation algorithms for
chemical-interested problems and could move this field significantly
forward. We also found that the precision of interested properties attained by this algorithm relies on the precision of implementing the time evolution of the molecular Hamiltonian in the experiment. Anyway, we made this early experimental progress
towards the long-term goal of a rapid quantum chemistry calculation
by quantum computers.

Note that, we have learned of a similar work\cite{Lanyon2009} done concurrently on photonic system since writing this paper.

The authors thank Kwek Leong-Chuan for help.
This work was supported by National Nature Science Foundation of
China, The CAS, Ministry of Education of P.R.China, the National
Fundamental Research Program.


\end{document}